\documentclass[12pt]{article}
\usepackage{graphicx}
\begin{document}
\title{Testing Einstein's time dilation under acceleration using M\"{o}ssbauer spectroscopy}
%
\author{Yaakov Friedman\thanks{Supported in part by German-Israel Foundation for Scientific Research and Development: GIF No. 1078-107.14/2009} and Israel Nowik\\
Jerusalem College of
Technology\\P.O.B. 16031 Jerusalem 91160, Israel\\email: friedman@jct.ac.il\\
Racah Institute of Physics, Hebrew University\\ Jerusalem 91904, Israel
\\ \\ Physica Scripta \textbf{85} 065702}

\maketitle

\begin{abstract}
The Einstein time dilation formula was tested in several experiments. Many trials have been made to measure the transverse second order Doppler shift by M\"{o}ssbauer spectroscopy using a rotating absorber, to test the validity of this formula. Such experiments are also able to test if the time dilation depends only on the velocity of the absorber, as assumed by Einstein's clock hypothesis, or the present centripetal acceleration contributes to the time dilation.
We show here that the fact that the experiment requires $\gamma$-ray emission and detection slits of finite size, the absorption line is broadened; by geometric longitudinal first order Doppler shifts immensely. Moreover, the absorption line is non-Lorenzian. We obtain an explicit expression for the absorption line for any angular velocity of the absorber.
 The analysis of the experimental results, in all previous experiments which did not observe the full absorption line itself, were wrong and the conclusions doubtful. The only proper experiment was done by K\"{u}ndig (Phys. Rev. 129 (1963) 2371), who observed the broadening, but associated it to random vibrations of the absorber.
We establish  necessary conditions for the successful measurement of a transverse second order Doppler shift by M\"{o}ssbauer spectroscopy.  We indicate how the results of such an experiment can be used to verify the existence of a Doppler shift due to acceleration and to test the validity of Einstein's clock hypothesis.

 \textit{PACS}: 76.80.+y, 03.30.+p

\textit{Keywords}: M\"{o}ssbauer effect; Absorption line; Time dilation experiments;
 Transverse Doppler shift; Einstein's clock hypothesis
\end{abstract}


\section{Introduction}

After the discovery of the M\"{o}ssbauer effect in 1958,  quantitative measurements of relativistic time dilation were carried out in the 1960s based on this effect  \cite{Hay1}-\cite{Kundig}, and the interest in such measurements lasts to this day \cite{Khoimetski},\cite{Khoimetski2}. The experiments \cite{Hay1}-\cite{Kundig} reported full agreement with the time dilation predicted by Einstein's theory of relativity. In the experiments \cite{Hay1}-\cite{Champeney2}, the M\"{o}ssbauer source was placed at the center of a fast rotating disk and an absorber at the rim of the disk. In the analyses of these experiments, it was assumed that the absorption line of the rotating absorber stays Lorentzian with same width as at rest, and is shifted only by the time dilation factor.

Based on the generalized principle of relativity and the ensuing symmetry,  in \cite{FG4}
it was shown that there are only two possible types of transformations between uniformly
accelerated systems. The validity of the \textit{Clock Hypothesis} is crucial to
determining which one of the two types of transformations is obtained. The Clock Hypothesis, as stated in
\cite{Einstein}, maintains that the rate of an accelerated clock is equal to that of a co-moving
unaccelerated clock.

If the Clock Hypothesis is not true, then the transformation is of Lorentz-type and implies the existence of
a universal maximal acceleration $a_m$ and an additional time dilation due to the
 acceleration of the clock. By \emph{acceleration}, we mean the proper acceleration defined \cite{Rindler} as $\mathbf{g}=d^2\mathbf{x}/dtd\tau$, where $\tau$ is the proper time.  In this case, it was shown in \cite{F09Ann} that a Doppler type shift for an accelerated source will be observed. This Doppler type shift is similar to the  Doppler shift due to the velocity of the source. The formulas for this shift are the same as those for the velocity Doppler shift, with $v/c$ replaced by $a/a_m$.

Consider an absorber placed on a disk, rotating with angular velocity $\omega$ at the distance $R$ from the center.  If this absorber is exposed to radiation of frequency $\nu_0$ through the center of the disk, then, since the velocity of the absorber is perpendicular to the radiation direction, the radiation will undergo a transverse Doppler shift $\nu_0(v^2/2c^2)=\nu_0(\omega ^2 R^2/2c^2)$ due to the time dilation of the absorber. If the conjecture about the existence of a shift due to acceleration is true, there will be an additional shift $\nu_0(a/a_m)=\nu_0(\omega^2 R/a_m)$, which is longitudinal, since the acceleration is in the direction of the radiation.

K\"{u}ndig's experiment \cite{Kundig} measured the transverse Doppler shift for a rotating disk by means
 of the M\"{o}ssbauer effect. In (only) this experiment, the absorption line of the rotating absorber was obtained. This experiment, as reanalyzed  by Kholmetskii \textit{et al} \cite{Khoimetski}, showed
 a significant deviation of about $20\%$ of the shift observed from the one predicted by special relativity. This deviation was explained in  \cite{F09Ann}  by the additional shift due to the acceleration. Moreover, the value of the maximal acceleration $a_m$ was estimated to be about  $10^{19}m/s^2$. This value of the maximal acceleration implies that the ratio of the new shift (due to acceleration) to the transversal one due to the velocity is
  \[\frac{a/a_m}{v^2/2c^2}= \frac{2c^2}{Ra_m}\approx 0.018/R\,,\]
 where $R$ is the distance ($R=9.2cm$ in \cite{Kundig}) of the absorber from the center of the disk.  K\"{u}ndig's experiment was not  designed to test the acceleration shift, and the recalculation by Kholmetskii \textit{et al} is not direct. Thus, the corrected K\"{u}ndig's result may serve only as an indication of the existence of a maximal acceleration and an estimate of its value.

  Note that in other time dilation experiments, the value of $R$  was about $5 \;cm$, implying that the ratio of the new shift to the known one is $35-40\%$. How was such a deviation not observed in \cite{Hay1}-\cite{Champeney2}? The reason for this is that their analyses did not take into account the broadening of the absorption curves during the rotation, as is explained below.

  We show here that the absorption line of a rotating M\"{o}ssbauer absorber gets broader during the
 rotation, even in the case when the absorber moves in the perpendicular direction to the radiation. This is due to the fact that the slit for the radiation beam leaving the source and the opening slit of the detector must have a finite width. Therefore, the velocity of the absorber is not perpendicular to all individual rays. Hence, these rays undergo a longitudinal Doppler shift in addition to the expected transversal one. This shift is very significant even for very small slits for the radiation beam and for detector, and leads to broadened non-Lorentzian absorption lines.

 Consider first the case when the absorber moves in the perpendicular direction to the radiation beam, as in Figure 1.
 \begin{figure}[h!]
  \centering
\scalebox{0.7}{\includegraphics{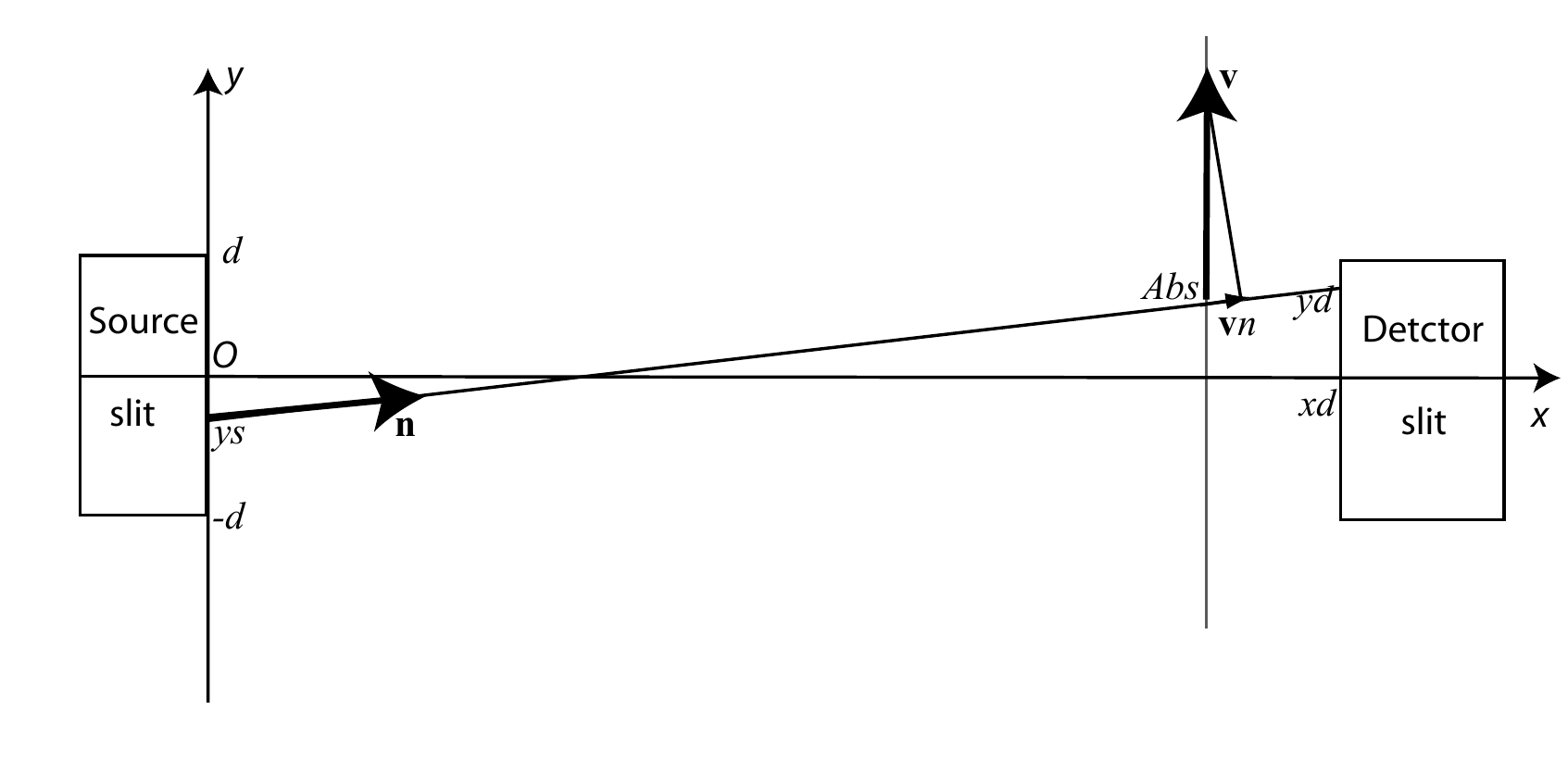}}
  \caption{Transverse Doppler shift with  absorber  moving perpendicular to the radiation beam.}
\end{figure}

Choose the direction of the radiation beam to be in the direction of the $x$-axis, and place the origin at center of the source and the detector at $L=x_d$. Denote by $\mathbf{v}$ the velocity of the absorber. In order for the transversal Doppler shift $v^2/2c^2$ to be observed, the velocity must be at least $100m/s.$ A ray leaves the source at the point $(0,y_s)$ and enters the detector at a point $(x_d,y_d)$. The  component of the velocity of the absorber in the direction $\mathbf{n}$ of the ray, for $d/L\ll 1$,  is $v_n=v(y_d-y_s)/L$, which may be significant even for very small width of the slits  (2d in Figure 1). This implies that most rays will also be exposed to a longtitudal Doppler shift and will change completely the shape of the M\"{o}ssbauer absorption line. One should observe a decrease in the absorption amplitude and a non-Lorentzian broadening of the absorption spectrum.

 To analyze the shifts that a ray undergoes in a rotating experiment, consider Figure 2.
 \begin{figure}[h!]
  \centering
\scalebox{0.8}{\includegraphics{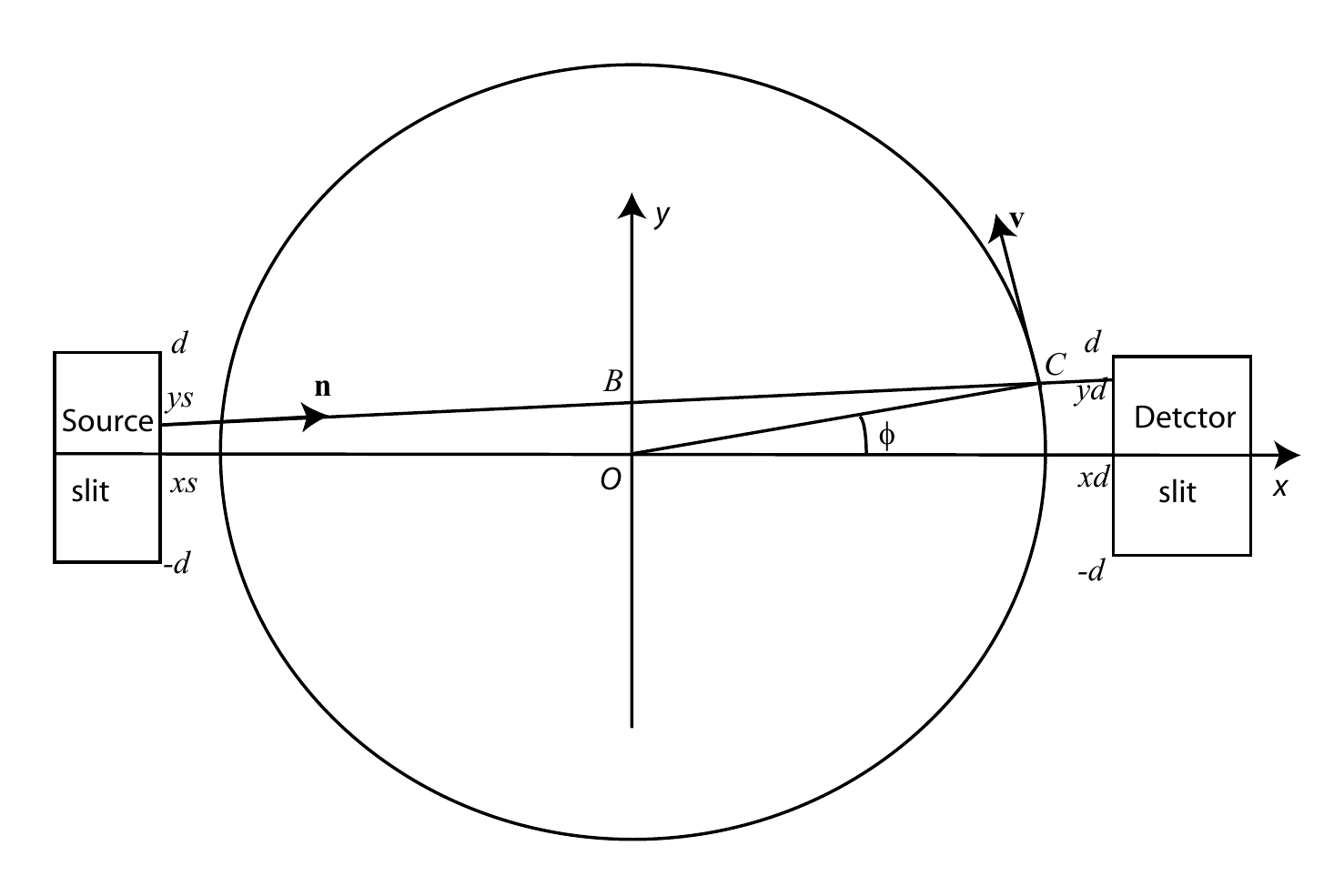}}
  \caption{Doppler shift with a rotating absorber.}
\end{figure}

 We choose again the direction of the radiation beam to be in the direction of the $x$-axis. We place the origin at the center of the rotating disk ($2R<L=x_d-x_s$). The radiation leaves the source at the point $(x_s,y_s)$ and enters the detector at a point $(x_d,y_d)$. The equation of the radiation line is
 \[y(x_d-x_s)+x(y_s-y_d)=y_sx_d-y_dx_s\,\]
and has the direction $\mathbf{n}=(x_d-x_s,y_d-y_s)/\sqrt{(x_d-x_s)^2+(y_d-y_s)^2}$, and intersects the $y$-axis at the point $B=(0,b)$, where
 \[b=\frac{y_sx_d-y_dx_s}{x_d-x_s},\]
 and $b$ obviously may have any value between $-d$ and $d$.
 Denote the distance of the absorber to the center by $R$, and denote the intersection point
 of our ray with the absorber by $C=(R\cos\phi,R\sin\phi)$. Since this point is on our radiation
 line, it satisfies the equation
 \begin{equation}\label{absorbPos}
   (x_d-x_s)R\sin\phi+(y_s-y_d)R\cos\phi=y_dx_s-y_sx_d\,.
 \end{equation}

 The velocity of the absorber at $C$ is $\mathbf{v}=R\omega(-\sin\phi,\cos\phi)$, where $\omega$ is
 the angular velocity of the disk. Thus, from equation (\ref{absorbPos}),
 we get that the component of the velocity of the absorber in the direction of the ray is
 \begin{equation}\label{vn}
   v_n= \mathbf{v}\cdot \mathbf{n}=\omega\frac{y_dx_s-y_sx_d}{\sqrt{(x_d-x_s)^2+(y_d-y_s)^2}}
   =-\omega b\frac{1}{\sqrt{1+(y_d-y_s)^2/(x_d-x_s)^2}}\,,
 \end{equation}
 which,  for $d/L\ll 1$, is approximately equal to $v_n=-\omega b$, and even then the velocity $v_n$ is extremely large for $\omega >10 s^{-1}$ and $b \approx 1 mm$, in comparison to the natural full width at half absorption intensity(
 $\Gamma=0.2 mm/s$) of the $^{57}Fe$  M\"{o}ssbauer absorption single line spectrum.

 A typical absorption curve is obtained by placing the source  on a transducer and moving it with a velocity $\mathbf{v}_s$ in the direction of the $x$-axis. This velocity is of order up to several millimeters per second. For such velocities, we may assume that the source velocity in the radiation  direction $\mathbf{n}$ is approximately equal $v_s$. The absorption curve $A_\omega(v_s)$ of an absorber rotating with angular velocity $\omega$ describes the rate of the radiation absorbed as a function of the velocity $v_s$ of the source. For a thin absorber with no chemical shift, a typical absorption curve for an absorber at rest  is a Lorentzian function $A_0(v_s)=a\gamma^2/(v_s^2+\gamma^2)$, of half width $\gamma$, absorption amplitude $a$, and resonance at $v_s=0$, as the one shown in Figure 3a for $\gamma=0.125mm/s$.

  For a rotating absorber, the total Doppler shift that the ray undergoes consists of a longitudinal shift due to the relative velocity of the source and the absorber and a transversal shift due to the time dilation of the absorber. All together are equivalent to a longitudinal shift of a source with velocity
 \begin{equation}\label{tildets}
   \tilde{v}_s=v_{s}-v_n+\frac{v^2}{2c}.
 \end{equation}
 Since the spread in $v_n$ is increasing with angular velocity, it contributes significantly to the spread of the observed absorption M\"{o}ssbauer line.

 Using (\ref{vn}) and (\ref{tildets}), the absorption curve of the rotating absorber will be
 \begin{equation}\label{AbsorRotatingAbs}
    A_\omega(v_s)=\int\int A_0(\tilde{v}_s)dy_sdy_d=\int\int A_0(v_{s}-\omega\frac{y_dx_s-y_sx_d}{\sqrt{(x_d-x_s)^2+(y_d-y_s)^2}}+\frac{R^2\omega ^2}{2c})dy_sdy_d\,,
 \end{equation}
where the integral has to be taken over the source emission intensity at $y_s$ and absorption by detector at $y_d$.
 This curve can be calculated analytically, if we assume  that the distributions along $y_s$ and $y_d$ are uniform on interval $[-d,d]$, $d/L\ll 1$ and $x_s=-x_d$. Under these assumptions $b=(y_s+y_d)/2$. Denoting $u=v_{s}+R^2\omega ^2/2c$ and defining a function $f(u)$ such that $f''(u)=A_0 (u)$, we get
 \[ A_\omega (u)=\frac{1}{d^2}\left(\int _{-d}^0 f''(u+\omega b)(d+b)db+
\int _0^d f''(u+\omega b)(d-b)db\right)\]\[=\frac{1}{d^2\omega}\left(f'(u+\omega b)(d+b)|_{-d}^0-\int _{-d}^0 f'(u+\omega b)db+f'(u+\omega b)(d-b)|_0^d +
\int _0^d f'(u+\omega b)db\right).\]
Thus,
\begin{equation}\label{Aomega}
   A_\omega (u)=\frac{f(u-\omega d)-2f(u)+f(u+\omega d)}{\omega ^2 d^2}.
\end{equation}

 If $d\mapsto 0$ and  $A_\omega (u)=af''(u)$, implying that the absorption curve of the rotating absorber will be $A_\omega(v_s)=A_0(v_s+\frac{R^2\omega ^2}{2c})$, which is a shift of the absorption curve of the non-rotating absorber. This was assumed in the experiments \cite{Hay1}-\cite{Champeney2}.

It is obvious that $A_\omega (u)$ is symmetric and $ A_\omega (v_s)$ is symmetric with respect to $v_s=-R^2\omega ^2/2c$. The results of the calculated absorption spectra  using equation (\ref{AbsorRotatingAbs}), using numerical integration, or using the analytic function (\ref{Aomega}) yield the same results, shown as transmission rate spectra line $T_\omega =1-A_\omega$  in Figure 3.
  \begin{figure}[h!]
  \centering
\scalebox{1}{\includegraphics{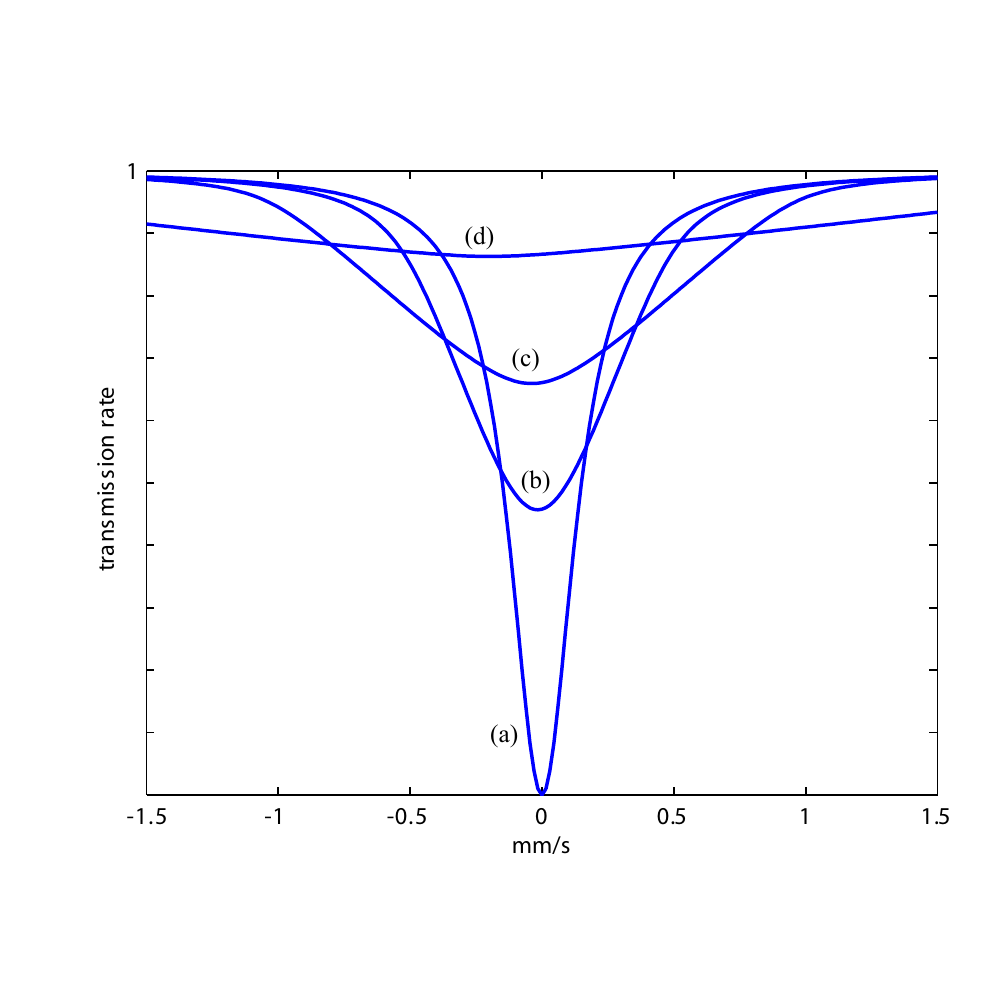}}
  \centering
  \caption{Transmission line $T_\omega(v_s)$ for $d=0.004\;mm$ and (a) $\omega=0$, (b) $\omega=500s^{-1}$, (c) $\omega=1000s^{-1}$, (d)  $\omega=3000s^{-1}$}
\end{figure}

The simulations show that even for small $d $, if $\omega$ becomes large enough  the absorption line  broadens drastically and even becomes unobservable. These curves have the same type of widening and shift as the absorption curves obtained experimentally in \cite{Kundig}.

In the experiments  \cite{Hay1}-\cite{Champeney2},\cite{Khoimetski2}, the value of $T(\tilde{v}_s)$ was measured only in the case in which the source and the absorber were relatively static, corresponding to $v_s=0$. Obviously in their experiments  $d>0$, (they do not specify), and their analysis did not consider the changes in shape of the absorption line as a function of $\omega$. Thus we cannot rely on their conclusions.

If the distributions of source emission intensity is $y_s$ dependent and the detection is $y_d$ dependent, but symmetric in the interval  $[-d,d]$, one will still observe  under the rotation a symmetric broadening.  If the distribution is not symmetric and has non-zero average, we will get an additional shift, but such a shift can be compensated for by reversing the direction of the rotation.

\noindent \textbf{Conclusions:} In order to measure the transverse Doppler shift by conventional M\"{o}sbauer spectroscopy with source mounted on a static transducer and the absorber moving on a fast rotating disc, one should take care of the following:
\begin{description}
  \item[(a)] One must measure the full spectrum of the absorption line.
   \item[(b)] One must measure this spectrum in both angular directions of the absorber. But most important;
    \item[(c)] One must put collimators on source and detector to reduce the slits to minimal dimensions, which will still allow a spectrum measurement in a reasonable time span.
                                                                               \end{description}

Only the experiment of W. K\"{u}ndig \cite{Kundig} obeyed at least (a) and (b).
    In his experiment he used a source of $^{57}Co:Fe$ against a $^{57}Fe$ enriched iron absorber. Thus both were almost identical and in both the iron nuclei were exposed to a strong magnetic field (33T) causing the emission and absorption spectra to be composed of six line patterns, which lead to observed multiline spectra as shown in Fig.4 (bottom), for various rotational frequencies of the absorber. Since K\"{u}ndig scanned the spectra in the range of about -1 to +1 mm/s, he observed only the central line with reduced intensity and line broadening, as the lines simulated in Fig.4 (middle and top).
 \begin{figure}[h!]
  \centering
\scalebox{0.8}{\includegraphics{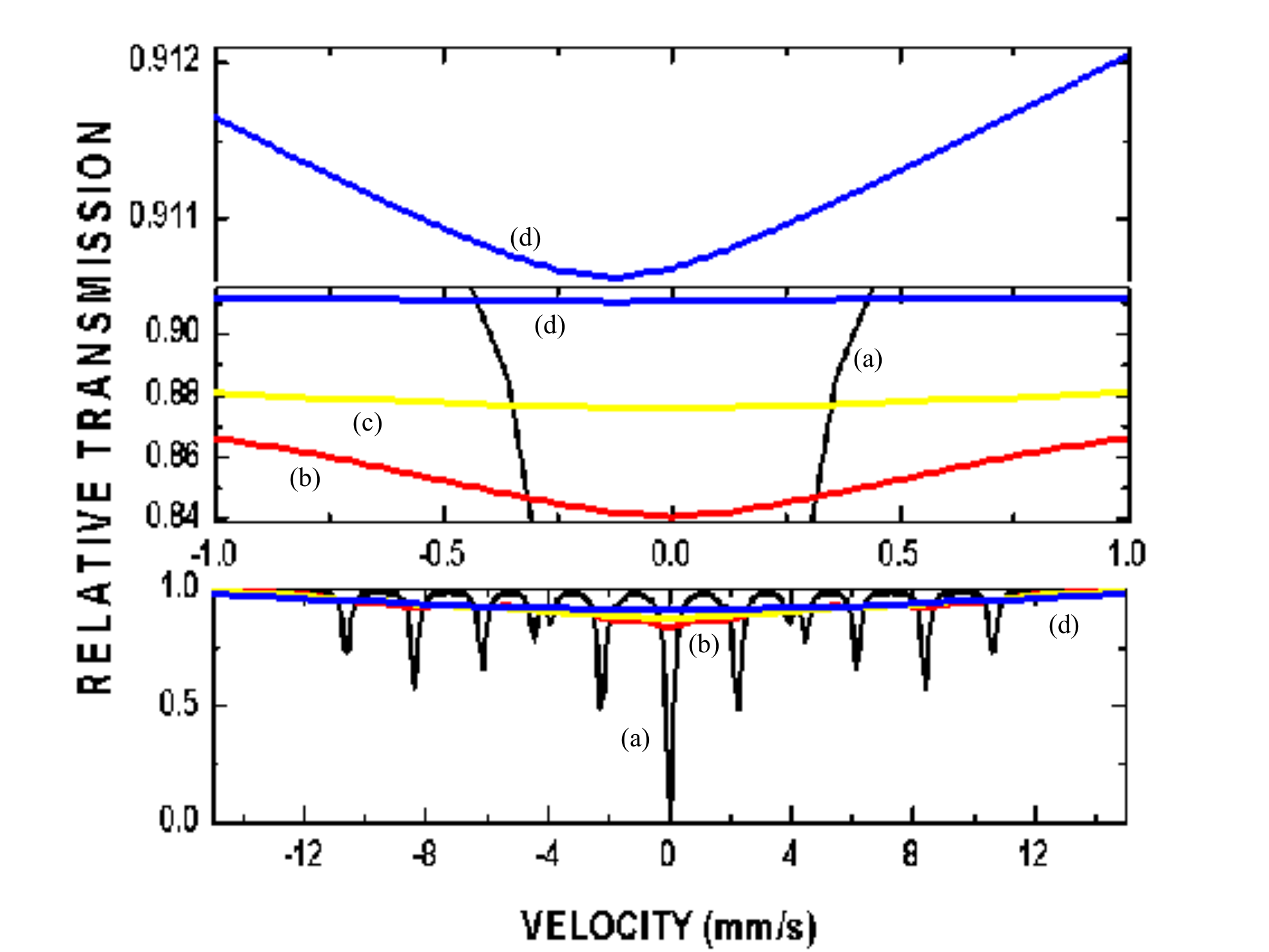}}
  \centering
  \caption{Simulated transmission spectra of a Co:Fe source and a Fe metal absorber, for a slit of  0.01 mm, and several rotational frequencies: (a) $\nu=1$, (b) $\nu=100$, (c) $\nu=200$, (d)  $\nu=500$ $s^{-1}$. The spectra resemble those experimentally observed by K\"{u}ndig. One observes clearly the second order Doppler shift in (d).}
\end{figure}

   Thus he actually observed the line broadening due to the finite slits, but interpreted them as due to rotor vibrations affecting the absorber. Nevertheless this experiment
 seems to be still the most accurate transverse Doppler shift experiment using the $^{57}Fe$ M\"{o}ssbauer effect.
His results with the corrected analysis reported in \cite{Khoimetski} indicates that the observed shift in this experiment is larger and deviates $\approx 20\%$ from that expected by special relativity theory.

 Einstein time dilation formula was verified by several experiments \cite{Baily77}-\cite{La"mmerzahl}. In W. K\"{u}ndig experiment the absorber was exposed to a significant acceleration. In \cite{F09Ann} it was indicated that the deviation of the shift in the experiment may be due to a longitudinal Doppler shift caused by the acceleration of the absorber. It led to an estimate of a universal maximal acceleration of order $10^{19}m/s^2$ and an indication of non-validity of the Einstein Clock Hypothesis \cite{Einstein}. For a long time, B. Mashhoon argued against the clock hypothesis and developed nonlocal transformations for accelerated observers (see the review article \cite{Mashhoon}).

  Note that time dilation M\"{o}ssbauer effect experiments have an advantage in identifying relatively small shift due to acceleration, since in these experiments the shift due to the velocity is of second order while the shift due the acceleration is of first order.
  Thus we recommend  M\"{o}ssbauer spectroscopy scientists to perform an accurate experiment measuring the shift of the absorption line for a fast rotating absorber which may reveal a new fundamental law \cite{F09Ann}, with a monumental effect on whole of physics.

 The authors want to thank  Eng. Menachem Friedman for the simulation of the results presented in Figure 3 and Dr. Tzvi Scarr for editorial proof.

\end{document}